\documentclass[epj]{svjour}
\usepackage{graphicx}
\usepackage{amsmath}

\journalname{Eur. Phys. J. A}
\usepackage{amssymb}
\newcommand{\be}{\begin{eqnarray}}
\newcommand{\ee}{\end{eqnarray}}
\newcommand{\bfq}{{\bf q}_{\perp}}

\newcommand{\bfk}{{\bf k}_{\perp}}

\usepackage{color}

\begin{document}
\title{Nucleon to $\Delta$ transition form factors and empirical transverse charge densities}
\author{Dipankar Chakrabarti \and Chandan Mondal}

\institute{Department of Physics, Indian Institute of Technology Kanpur, Kanpur-208016, India.}

\date{Received: date / Revised version: date}

\abstract{We investigate the nucleon to $\Delta$ transition form factors  in a soft-wall AdS/QCD model and a light-front quark-diquark model inspired by AdS/QCD. From the transition form factors we evaluate the transition charge densities which influences the nucleon to $\Delta$ excitation. Here we consider both the unpolarized and the transversely polarized cases. The AdS/QCD predictions are compared with available experimental data and with the results of the global parameterization, MAID2007.
\PACS{
  {12.38.-t}{Quantum chromodynamics} \and
  {13.40.Gp}{Electromagnetic form factors} \and
  {14.20.-c}{Baryons}\and
  {21.10.Ft}{Charge distribution}
} 
} 
\authorrunning{D. Chakrabarti\and C. Mondal}
\titlerunning{Nucleon to $\Delta$ transition form factors and empirical transverse charge densities}
\maketitle
\vskip0.2in
\noindent
\section{ Introduction}
The excitation of nucleon resonances in electromagnetic interactions plays a crucial role in the physics of the strong interaction. 
There have been considerable efforts to   investigate the nucleon to $\Delta(1232)$ resonance both  experimentally and theoretically \cite{Bermuth:1988ms,Fiolhais:1996bp,Lu:1996rj,Faessler:2006ky,Pascalutsa:2005ts,Kamalov:1999hs,Kamalov:2000en,Sato:2000jf,tiator1,Matsuyama:2006rp}. For detailed review on this subject, we refer to the articles \cite{Pascalutsa:2006up,tiator2} and the references therein. 
The form factors which describe the transition of the nucleon to its first excited state, $\Delta$(1232) provide us the information about the sensitivity on the nucleon shape \cite{Pascalutsa:2006up,Bernstein:2007jt}. It is always very difficult to calculate the hadron properties such as the mass spectrum, form factors, and parton distributions directly from the first principles of QCD, because they require non-perturbative methods. The chiral effective field theory is able to provide $N\rightarrow \Delta$ from factors for the low momentum transfer \cite{Bermuth:1988ms,Fiolhais:1996bp,Lu:1996rj,Faessler:2006ky,Pascalutsa:2005ts}. 
Large $N_c$ limit gives a simplified picture of QCD but provides us many interesting insights and a good approximation of real QCD. 
There are some applications of the large $N_c$ limit which leads to the understanding of electromagnetic transitions between the $\gamma^*N \rightarrow \gamma\Delta(1232)$ \cite{Pascalutsa:2006up,pascal,Grigoryan:2009pp,frank} and the predictions are made for a wide range $Q^2$, for example, the magnetic-dipole transition amplitude can be extended up to $Q^2\sim 6-8~\rm GeV^2$. Using the relation between the isovector nucleon magnetic moment and the $N\rightarrow \Delta$ transition magnetic moment as shown in \cite{Jenkins:1994md}, Pascalutsa and Vanderhaeghen \cite{pascal} have established that in the large-$N_c$ limit the $N\rightarrow \Delta$ transition form factors can be expressed in terms of nucleon electromagnetic form factors.  The consistency of large-$N_c$ relations for a finite momentum transfer have been verified in an empirical parameterization of the nucleon form factors \cite{Brad} by the comparison of transition ratios with the  experimental data. 
It is therefore interesting
to investigate whether the large-$N_c$ relations proposed by Pascalutsa and Vanderhaeghen are valid for some other phenomenological models such as AdS/QCD or quark models. Considerable progress has been achieved as well in the lattice QCD simulations of nucleon to $\Delta$ transition. The transition form factors have been evaluated by lattice QCD calculated in \cite{Alexandrou:2007dt}. One of the most successful phenomenological model for the transition form factors is provided by the Mainz unitary isobar model (MAID) for the pion photo and electroproduction on the nucleon,  the $Q^2$ dependence of the parameters for  transition form factors is presented in MAID2007 \cite{maid} which agrees well with  the CLAS data from JLab for the magnetic-dipole $N \rightarrow \Delta(1232)$ amplitude and the ratios of transition form factors. Recently, a new set of empirical parameterizations for $N\to \Delta$ transition amplitude has been proposed in \cite{Ramalho:2016zzo}.

Form factors of nucleon reflect the special distributions such as charge densities via a Fourier transform. The form factors involve initial and final  states with different momenta and three dimensional Fourier transforms can not be interpreted as densities. But the transverse densities defined in fixed light-front time are free from this difficulty and have proper density interpretation \cite{miller07,miller10,miller09,venkat}. Similarly, the empirical knowledge of transition form factors in a wide range of $Q^2$ also allows one to map out the quark transverse charge distributions that induce the transitions \cite{tiator2,Tiator:2008kd,Tiator:2009mt,vande}.  In this work, we evaluate the $N\to \Delta$ transition form factors and the transition charge distributions in the AdS/QCD models.

Recently, AdS/QCD correspondence has achieved significant attraction in research of non-perturbative QCD because this formalism has been proven as one of the most promising techniques to investigate the structure of hadron. The AdS/CFT duality, also known as the Maldacena conjecture \cite{maldacena} relates a strongly coupled gauge theory in $d$ space-time dimensions by a dual weak coupling gravity theory in AdS$_{d+1}$ space.  There are many applications of AdS/CFT duality to investigate the QCD phenomena \cite{PS1,PS2,costa1,costa2,costa3,costa4}. Since QCD is not a conformal theory, to apply AdS/CFT to QCD, the conformal invariance need to be broken. In the literature, there are many efforts to break the conformal symmetry in the gravity side \cite{Sakai:2004cn,Sakai:2005yt,Gubser:2008ny}. In light-front holography, two models are adopted to achieve this goal, one is called hard wall model where one puts a boundary in the AdS space so that the wave functions are made to vanish at he boundary and the other is called the soft wall 
model where a confining potential is introduced in the AdS space which breaks the conformal invariance and generates the mass spectrum. For the 
baryon sector, the AdS/QCD formalism has been developed by  several groups \cite{BT00,BT011,BT012,katz,SS,AC,ads11,ads12,ModelII,ads21,ads22,ads23}. Although this correspondence gives only the semi-classical approximation of QCD, so far the framework has  been successfully applied to  describe many hadron properties such as hadron mass spectrum, parton distribution functions, GPDs, meson and nucleon form factors, transverse densities, structure functions etc. \cite{AC,ads11,ads12,ModelII,ads21,ads22,ads23,AC4,BT11,BT12,BT1b1,BT1b2,BT2,deTeramond:2013it,Rad,branz1,branz2,vega1,vega2,vega3,CM,CM2,CM3,Mondal:2015fok,HSS,abidin08,BT_new3}. The applications in AdS/QCD to nucleon resonances have been studied in \cite{BT2,BT_new3,reso}. The form factors of $\Delta$  baryons (spin $3/2$) and the $N\rightarrow \Delta$ transition form factors in the AdS/QCD framework have been reported in \cite{hong2}.
Recently, a light-front quark-diquark model for nucleon has been developed in \cite{Gut} where the wavefunctions are constructed from the soft-wall AdS/QCD and this has been extensively used to investigate many interesting properties of the nucleons \cite{CM41,CM5,CM6,CM7,Chakrabarti:2016yuw,CM42,Chakrabarti:2016mwn}. In this work, we study the nucleon to $\Delta(1232)$ transition form factors in a soft-wall AdS/QCD model as well as a light-front quark-diquark model inspired by AdS/QCD using the large-$N_c$ relations established by Pascalutsa and Vanderhaeghen \cite{pascal}. The transition charge densities which influences the nucleon to $\Delta$ excitation are also investigated by taking two-dimensional Fourier transforms of the transition form factors. We compare our results with the available experimental data as well as with the global parameterization, MAID2007 \cite{maid}. 

The paper is organized as follows. The electromagnetic form factors for nucleon in the soft-wall AdS/QCD and the quark-diquark models have been given in Sec.\ref{ads}. We present the nucleon to $\Delta(1232)$ transition form factors in the  Sec.\ref{ND_FFs}. In Sec.\ref{ND_densities}, the empirical transverse charge densities in the nucleon to $\Delta$ excitation for both unpolarized and transversely polarized cases have been discussed. Then we provide a brief summary in Sec.\ref{concl}.
\section{Nucleon form factors}\label{ads}
\subsection{Soft-wall AdS/QCD model}
We consider  the AdS/QCD model for nucleon form factors proposed by Brodsky and T\'{e}ramond \cite{BT2}.  Here, we describe the salient points of the model in brief.
The relevant AdS/QCD action in the soft-wall model for the fermion field is written as
\be
S&=&\int d^4x dz \sqrt{g}\Big( \frac{i}{2}\bar\Psi e^M_A\Gamma^AD_M\Psi -\frac{i}{2}(D_M\bar{\Psi})e^M_A\Gamma^A\Psi\nonumber\\
&&-\mu\bar{\Psi}\Psi-V(z)\bar{\Psi}\Psi\Big),\label{action}
\ee
where $e^M_A=(z/R)\delta^M_A$ is the inverse  vielbein and $V(z)$ is the confining potential which breaks the conformal invariance and 
 $R$ is the AdS radius. For $d=4$ dimensions, $\Gamma_A=\{\gamma_\mu, -i\gamma_5\}$. 
One can derive the Dirac equation in AdS from the above action:
\be
i\Big(z \eta^{MN}\Gamma_M\partial_N+\frac{d}{2}\Gamma_z\Big)\Psi -\mu R\Psi-RV(z)\Psi=0.\label{ads_DE}
\ee
It is possible to map the Dirac equation in AdS space with the light front wave equation, by identifying the holographic variable, $z\to\zeta$, where $\zeta$ is the light front transverse variable which measures the separation of the  quark and 
gluonic constituents in the hadron and substituting $\Psi(x,\zeta)=e^{-iP\cdot x}\zeta^2\psi(\zeta)u(P)$ in Eq.(\ref{ads_DE}) and setting $\mid \mu R\mid=\nu+1/2$ where  $\nu$ is related with the orbital angular momentum by $\nu=L+1$ .
For linear confining potential  $U(\zeta)=(R/\zeta)V(\zeta)=\kappa^2\zeta$, one obtains the light front wave equation for the baryon in $2\times 2$ spinor representation as
\be
\big(-\frac{d^2}{d\zeta^2}-\frac{1-4\nu^2}{4\zeta^2}+\kappa^4\zeta^2&+&2(\nu+1)\kappa^2\Big)\psi_+(\zeta)\nonumber\\&=&{\cal{M}}^2\psi_+(\zeta),\label{we1}\\
 \big(-\frac{d^2}{d\zeta^2}-\frac{1-4(\nu+1)^2}{4\zeta^2}&+&\kappa^4\zeta^2+2\nu\kappa^2\Big)\psi_-(\zeta)\nonumber\\&=&{\cal{M}}^2\psi_-(\zeta),\label{we2}
 \ee 
The specific form of the confining potential $U(\zeta)$ is chosen because of using the potential one can reproduce linear Regge trajectories for the baryon mass spectrum. Again, squaring the Dirac equation with $U(\zeta)$, one can generate a Klein-Gordon equation with the potential $\kappa^4 z^2$ which is consistence with the same confining potential appeared in the meson sector from the dilaton field~\cite{BT2}.
The wave Eqs.(\ref{we1}) and (\ref{we2}) lead to the AdS solutions of nucleon wavefunctions $\psi_+(z)$ and $\psi_-(z)$ corresponding to different orbital angular momentum $L^z=0$ and $L^z=+1$ combined with spin components $S^z = +1/2$ and $S^z =  - 1/2$ respectively \cite{BT2},
\be
\psi_+(\zeta)\sim\psi_+(z) &=& \frac{\sqrt{2}\kappa^2}{R^2}z^{7/2} e^{-\kappa^2 z^2/2}\label{psi+},\\
\psi_-(\zeta)\sim\psi_-(z) &=& \frac{\kappa^3}{R^2}z^{9/2} e^{-\kappa^2 z^2/2}\label{psi-}.
\ee 
$\psi_+(z)$ and $\psi_-(z)$ represent the $S$ and $P$ components of proton with equal probability \cite{deTeramond:2013it}. 
The eigenvalues of the wave Eqs.(\ref{we1}) and (\ref{we2}) are
\be \label{Mplus}
\mathcal{M}_+^2 &=& \left(4n + 2 \nu + 2\right ) |\kappa^2| + 2 \left( \nu  +1 \right) \kappa^2,\\
\mathcal{M}_-^2 &=& \left(4n + 2(\nu +1) +2 \right) |\kappa^2| + 2 \nu \kappa^2.
\ee
For $\kappa^2>0$ one finds $\mathcal{M}_+^2 = \mathcal{M}_-^2 = \mathcal{M}^2$ where
\be \label{M2F}
\mathcal{M}^2 = 4 \, \kappa^2  \left(  n +  \nu + 1 \right),
\ee
identical for both $\psi_+(\zeta)$ and $\psi_-(\zeta)$. For $\kappa^2<0$, it follows that $\mathcal{M}_+^2 \neq \mathcal{M}_-^2$ and no solution is possible~\cite{deTeramond:2013it,BT_new3}.

The Dirac form factors in this model are obtained by the SU(6) spin-flavor symmetry and given by~\cite{BT2,CM2,BT_new3}
\be
F_1^p(Q^2)&=&R^4\int \frac{dz}{z^4} V(Q^2,z)\psi^2_+(z),\label{F1p} \\
F_1^n(Q^2)&=& -\frac{1}{3}R^4\int \frac{dz}{z^4} V(Q^2,z)(\psi^2_+(z)-\psi^2_-(z)).\label{F1n}
\ee
Using the action in Eq.(\ref{action}), a precise mapping for the spin-flip nucleon form factor is not possible. Thus, the Pauli form factors for the nucleons are modeled  as~\cite{BT2,CM2,BT_new3}
\be
F_2^{p/n}(Q^2) =  \kappa_{p/n}R^4 \int \frac{dz}{z^3}\psi_+(z) V(Q^2,z) \psi_-(z).\label{F2}
\ee
The Pauli form factors are normalized to $F_2^{(p)n}(0) = \kappa_{(p)n}$ where $\kappa_{(p)n}$ are the anomalous magnetic moment of (proton)neutron. The bulk-to-boundary propagator, $V(Q^2,z)$ for soft wall model is given by \cite{BT2,Rad}
\be
\! V(Q^2,z)=\kappa^2z^2\int_0^1\!\frac{dx}{(1-x)^2} x^{Q^2/(4\kappa^2)} e^{-\kappa^2 z^2 x/(1-x)}.
 \ee
There is only one free parameter $\kappa$ in this soft-wall model. We use the value of scale parameter $\kappa=0.4~\rm GeV$ which is fixed by fitting the ratios of Pauli and Dirac form factors for proton with the experimental data \cite{CM,CM2}. 
\subsection{Light-front quark-diquark model}
In the light-cone formalism for a composite system of spin $\frac{1}{2}$, the Dirac and Pauli form factors $F_1(q^2)$ and $F_2(q^2)$ are related to the helicity-conserving and helicity-flip matrix elements of the $J^+$ current \cite{BD}:
\be
\langle P+q, \uparrow|\frac{J^+(0)}{2P^+}|P, \uparrow\rangle &=&F_1(q^2),\\
\langle P+q, \uparrow|\frac{J^+(0)}{2P^+}|P, \downarrow\rangle &=&-(q^1-iq^2)\frac{F_2(q^2)}{2M_n},
\ee
where $M_n$ is the nucleon mass. In quark-diquark model, the three valence quarks of nucleon are considered as an effectively composite system composed of a fermion (quark) and a composite  state of diquark based on one loop quantum fluctuations.
Here we consider a light front quark-diquark model for nucleon \cite{Gut} where the 2-particle wavefunction is modeled from the soft-wall AdS/QCD solution. In the light front quark-diquark model, writing nucleon as a two particle bound state of a quark and a diquark, one can evaluate the Dirac and Pauli form factors for quarks in terms of overlap of the wavefunctions \cite{BD,BHMI1,BHMI2,BHMI3} as
\be
F_1^q(Q^2) &=& \int_0^1dx \int \frac{d^2\bfk}{16\pi^3}~\Big[\psi_{+q}^{+*}(x,\bfk')\psi_{+q}^+(x,\bfk) \nonumber\\
&+&\psi_{-q}^{+*}(x,\bfk')\psi_{-q}^+(x,\bfk)\Big],\\
F_2^q(Q^2) &=& -\frac{2M_n}{q^1-iq^2}\int_0^1dx \int \frac{d^2\bfk}{16\pi^3}~\nonumber\\
&\times&\Big[\psi_{+q}^{+*}(x,\bfk')\psi_{+q}^-(x,\bfk)\nonumber\\
&+&\psi_{-q}^{+*}(x,\bfk')\psi_{-q}^-(x,\bfk)\Big],
\ee
where $\bfk'=\bfk+(1-x)\bfq$. $\psi_{\lambda_q q}^{\lambda_N}(x,\bfk)$ are the LFWFs with nucleon helicities $\lambda_N=\pm$ and for the struck quark $\lambda_q=\pm$, where plus and minus correspond to $+\frac{1}{2}$ and $-\frac{1}{2}$ respectively. We consider the frame, $q=(0,0,\bfq)$,  thus $Q^2=-q^2=\bfq^2$. The LFWFs are 
specified at an initial scale $\mu_0=313$~MeV \cite{Gut} :  
\be 
\psi_{+q}^+(x,\bfk) &=&  \varphi_q^{(1)}(x,\bfk) 
\,, \nonumber\\
\psi_{-q}^+(x,\bfk) &=& -\frac{k^1 + ik^2}{xM_N}   \, \varphi_q^{(2)}(x,\bfk) \,, \\
\psi_{+q}^-(x,\bfk) &=& \frac{k^1 - ik^2}{xM_N}  \, \varphi_q^{(2)}(x,\bfk)
\,, \nonumber\\
\psi_{-q}^-(x,\bfk) &=& \varphi_q^{(1)}(x,\bfk) \nonumber
\,, \label{WF}
\ee
where $\varphi_q^{(i)}(x,\bfk)~(i=1,2)$ is the wave functions predicted by soft-wall AdS/QCD, modified by introducing the tunable parameters $a_q^{(i)}$ and $b_q^{(i)}$ for quark $q$ \cite{Gut}:
\be
\varphi_q^{(i)}(x,\bfk)&=&N_q^{(i)}\frac{4\pi}{\kappa}\sqrt{\frac{\log(1/x)}{1-x}}x^{a_q^{(i)}}(1-x)^{b_q^{(i)}}\nonumber\\
&\times&\exp\bigg[-\frac{\bfk^2}{2\kappa^2}\frac{\log(1/x)}{(1-x)^2}\bigg].
\ee
$\varphi_q^{(i)}(x,\bfk)$ reduces to the AdS/QCD prediction for $a_q^{(i)}=b_q^{(i)}=0$ \cite{BT2}. The AdS/QCD scale parameter, $\kappa$ is taken to be $0.4$ GeV \cite{CM,CM2}. All the parameters $a^{(i)}_q$ and $b^{(i)}_q$ with the constants $N^{(i)}_q$  are fixed by fitting the electromagnetic properties of the nucleons \cite{CM6}. 
\begin{figure*}[htbp]
\begin{center}
\includegraphics[width=7.9cm,clip]{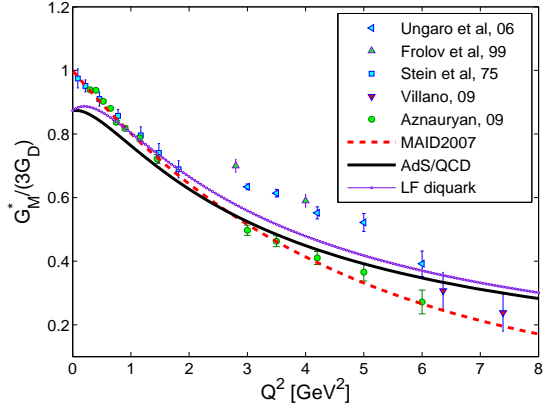}
\end{center}
\caption{\label{TFF}(Color online) The $N \to \Delta$ transition Jones-Scadron form factor $G_M^*(Q^2)/(3G_D)$, 
where the dipole form factor $G_D(Q^2)=1/(1+Q^2/0.71~\rm GeV^2)^2$. The red dashed line represents the global fit MAID2007 \cite{maid}; the solid black line and the purple line with circle represent the soft-wall AdS/QCD and the light-front quark-diquark models respectively. The experimental data are taken from Refs.\cite{stein,vilano,azn,ungaro,frolov}.}
\end{figure*}
\begin{figure*}[htbp]
\begin{minipage}[c]{0.98\textwidth}
{(a)}\includegraphics[width=7.9cm,clip]{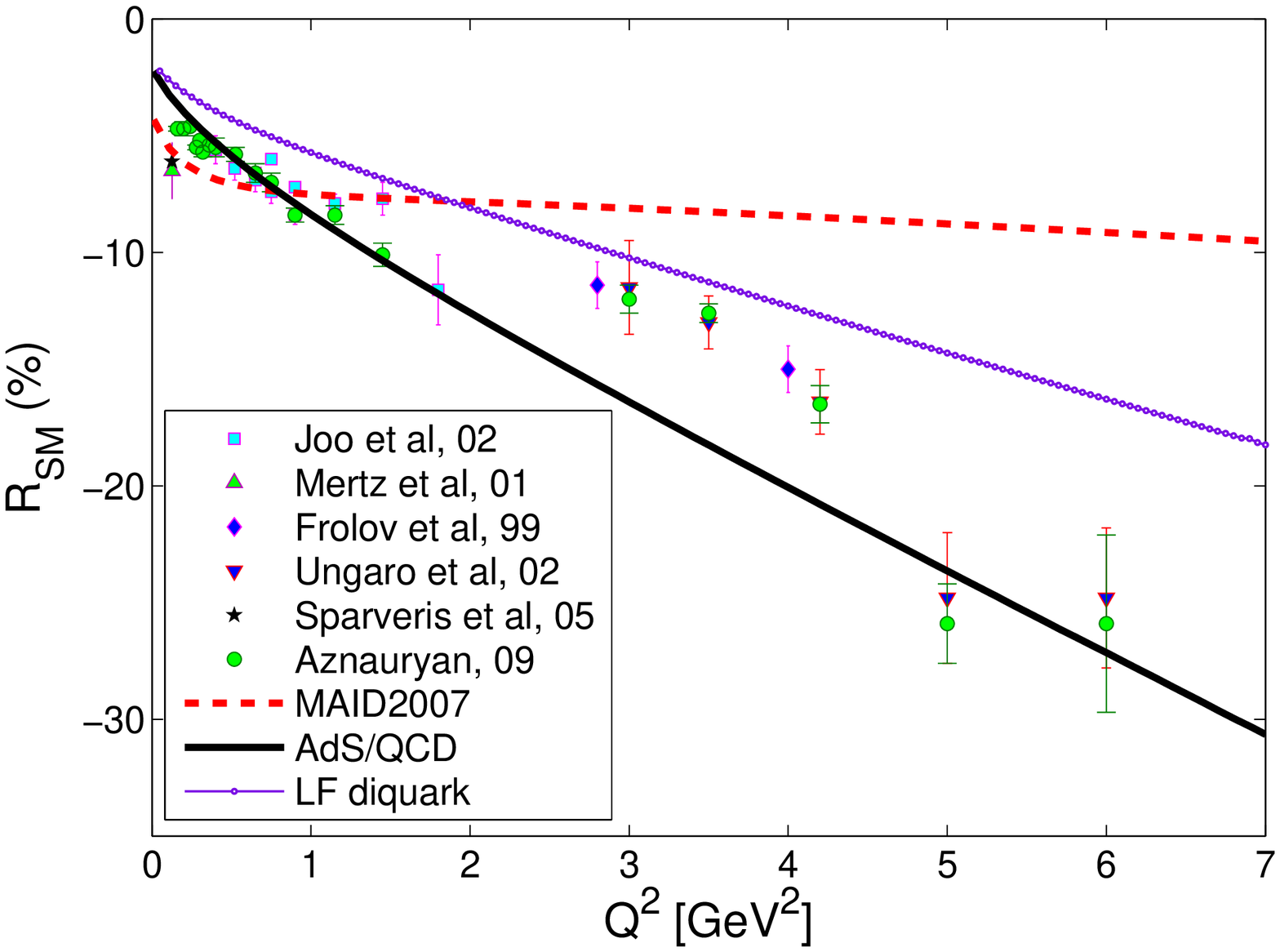}
{(b)}\includegraphics[width=7.9cm,clip]{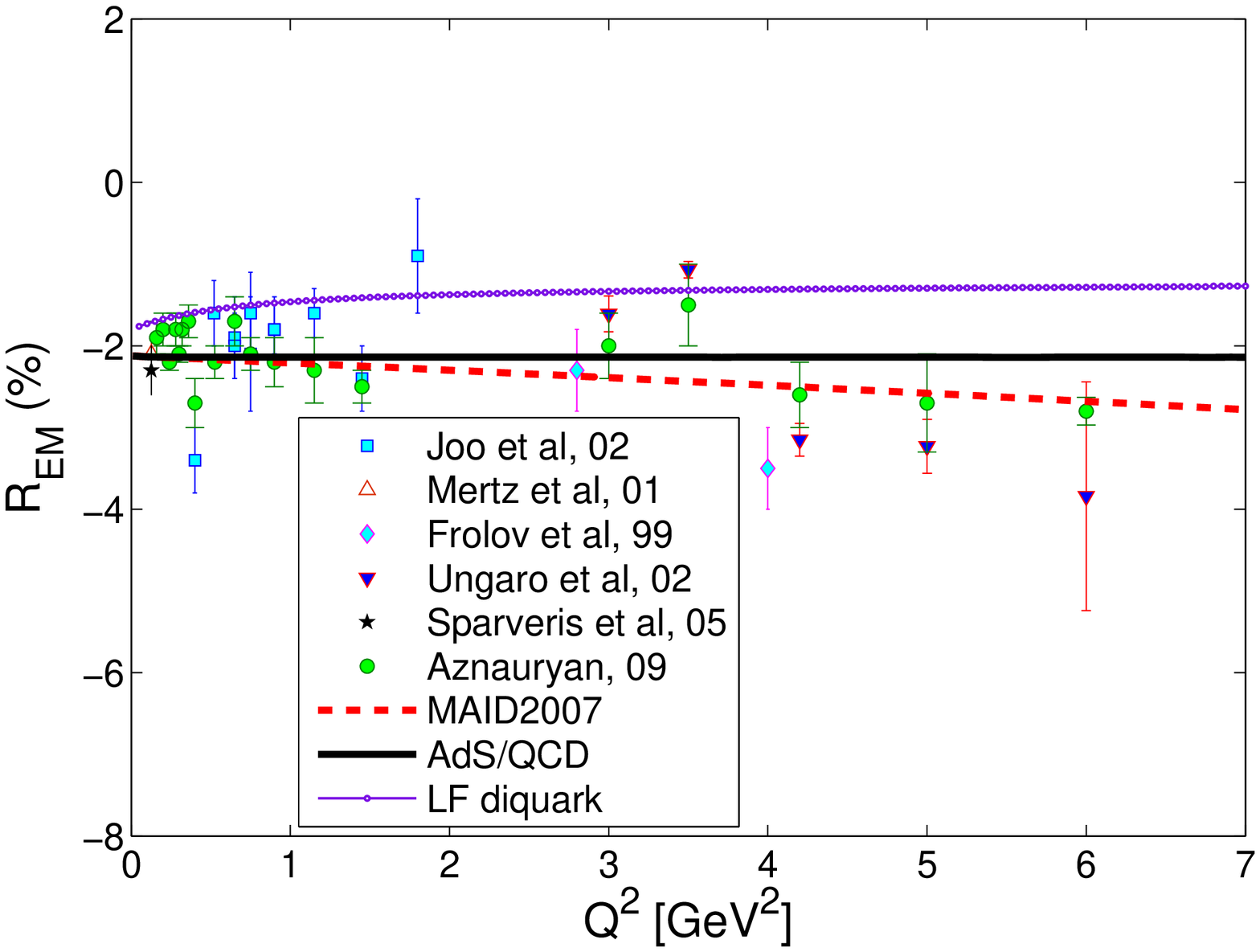}
\end{minipage}
\caption{\label{ratios}(Color online) The ratios of transition form factors (a) $R_{SM}(Q^2)$ and (b) $R_{EM}(Q^2)$. The red dashed line represents the global fit MAID2007 \cite{maid}. 
The experimental data are taken from Refs.\cite{azn,ungaro,frolov,joo,mertz,spar}.} 
\end{figure*}

\section{ $N\to \Delta$ transition form factors}\label{ND_FFs}
The electromagnetic nucleon to $\Delta$ transition form factors in the large-$N_c$ limit can be expressed entirely in terms of the nucleon
electromagnetic properties. For example, it has been established  that the magnetic nucleon to $\Delta$ transition moment is related to the isovector anomalous magnetic moment of the nucleon \cite{Jenkins:1994md}
\be
G_M^*(0)&=&\frac{1}{\sqrt{2}}\kappa_V,\label{Gm0}
\ee
where the isovector anomalous magnetic moment of the nucleon, $\kappa_V\simeq3.7$.
The relation in Eq.(\ref{Gm0}) has been extended for finite $Q^2$ in \cite{pascal,frank} as
\be
G_M^*(Q^2)=\frac{1}{\sqrt{2}}[F_2^p(Q^2)-F_2^n(Q^2)],\label{Gm}
\ee
where $F_2^p(Q^2)$ and $F_2^n(Q^2)$ are the Pauli form factors of proton and neutron. 
It has also been predicted that in the large-$N_c$ limit, the $G_{E,C}^*(Q^2)$ can be expressed in term of neutron electric form factor $G_E^n(Q^2)$ for finite but small $Q^2$ as \cite{pascal}
\be
G_E^*(Q^2)=\bigg(\frac{M_N}{M_{\Delta}}\bigg)^{3/2}\frac{M_{\Delta}^2-M_N^2}{2\sqrt{2}Q^2}G_E^n(Q^2),\label{Ge}
\ee
and
\be
G_C^*(Q^2)=\frac{4M_{\Delta}^2}{M_{\Delta}^2-M_N^2}G_E^*(Q^2),\label{Gc}
\ee
or equivalently the ratios
\be
R_{EM}(Q^2)=-\frac{G_E^*}{G_M^*},  \quad R_{SM}(Q^2)=-\frac{Q_+Q_-}{4M_{\Delta}^2}\frac{G_C^*}{G_M^*},\label{Eq_ratios}
\ee
where $Q_{\pm}=\sqrt{(M_{\Delta}\pm M_N)^2+Q^2}$ and $M_N(M_{\Delta})$ is the mass of nucleon($\Delta$). The electric form factors $G_E(Q^2)$ are expressed in terms of Dirac and Pauli form factors as
\be
G_E(Q^2)=F_1(Q^2)-\frac{Q^2}{4M_N}F_2(Q^2).
\ee
Putting all the relations of Eqs.(\ref{Gm}), (\ref{Ge}) and (\ref{Gc}) together in Eq.(\ref{Eq_ratios}) one obtains the following
expression for the ratios of nucleon to $\Delta$ transition form factors in terms of the nucleon form
factors: 
\be
R_{EM}(Q^2)&=&-\bigg(\frac{M_N}{M_{\Delta}}\bigg)^{3/2}\frac{M_{\Delta}^2-M_N^2}{2Q^2}\nonumber\\
&\times&\frac{G_E^n(Q^2)}{F_2^p(Q^2)-F_2^n(Q^2)},\label{Re}\\
R_{SM}(Q^2)&=&-\bigg(\frac{M_N}{M_{\Delta}}\bigg)^{3/2}\frac{Q_+Q_-}{2Q^2}\nonumber\\
&\times&\frac{G_E^n(Q^2)}{F_2^p(Q^2)-F_2^n(Q^2)}.\label{Rs}
\ee
Using the empirical parameterization of the nucleon FFs by Bradford $et~al$ \cite{Brad}, the above relations of nucleon to $\Delta$ transition ratios in Eqs.(\ref{Re}) and (\ref{Rs}) have been verified by comparing with the experimental data \cite{pascal}.

We evaluate the nucleon to $\Delta$ transition form factors using the nucleon electromagnetic form factors obtained in the framework of a soft-wall AdS/QCD model as well as in a light-front quark-diquark model.
In Fig.\ref{TFF}, we show the $Q^2$ dependence of nucleon to $\Delta$ transition form factor $G_M^*$ and the transition ratios $R_{SM}$ and $R_{EM}$ are shown in Fig.\ref{ratios}.
The results are compared with the available experimental data and also with the standard parameterization, MAID2007 \cite{maid}. For $G_M^*(Q^2)$, both the results of AdS/QCD and the quark-diquark model are in good agreement with the experimental data whereas for the transition ratios, AdS/QCD results are consistent with the experimental data but   quark-diquark model deviates a bit. Although the relations in Eqs.(\ref{Re}) and (\ref{Rs}) have been derived assuming that the momentum transfer is small, $Q^2< 1~\rm GeV^2$ \cite{pascal}, one can notice that the AdS/QCD prediction for the ratios is in well agreement with the experimental data at high $Q^2$ whereas the quark-diquark model as well as the MAID 2007 parameterization are unable to reproduce the data at large $Q^2$.
\section{ $N\to\Delta$ transition empirical charge densities}\label{ND_densities}
We study the quark transition charge densities in the transverse plane using the empirical information on the nucleon to $\Delta$ transition form factors which are characterized by the Jones-Scadron form factors $G_M^*$, $G_E^*$ and $G_C^*$ \cite{jones}. According to the standard interpretation \cite{miller07,miller10,miller09,venkat,Tiator:2008kd,Tiator:2009mt,vande,CM3,selyugin,weiss} the charge densities in the transverse plane can be identified with the two dimensional Fourier transform of the electromagnetic form factors in the light-cone frame when the momentum transfer is purely in transverse direction. One defines the quark transition charge densities in transverse plane in the following way 
\be
\rho_0^{N\Delta}(b)&=&\int \frac{d^2\Delta_{\perp}}{(2\pi)^2}e^{i\Delta_{\perp}. b_{\perp}}\frac{1}{2P^+}\nonumber\\
&\times&\Big\langle P^+,\frac{\Delta_{\perp}}{2},\lambda_{\Delta}|J^+(0)|P^+,-\frac{\Delta_{\perp}}{2},\lambda_{N}\Big\rangle, \label{defi}
\ee
where $\lambda_N$ and $\lambda_{\Delta}$ are the light-front helicities of nucleon and $\Delta(1232)$ and $\Delta_{\perp}=Q(\cos\phi_q \hat{x}+\sin\phi_q \hat{x})$. The matrix elements of the electromagnetic current $J^+(0)$ operator between nucleon and $\Delta$ relate the transition form factors as
\be
&&\Big\langle P^+,\frac{\Delta_{\perp}}{2},\lambda_{\Delta}|J^+(0)|P^+,-\frac{\Delta_{\perp}}{2},\lambda_{N}\Big\rangle\nonumber\\
&=&(2P^+)e^{i(\lambda_N-\lambda_{\Delta})\phi_q}G^+_{\lambda_N\lambda_{\Delta}}.\label{rg}
\ee
Here $G^+_{\lambda_N\lambda_{\Delta}}$ are the transition form factors which can equivalently be written in terms of $G_M^*$, $G_E^*$ and $G_C^*$. 
Thus, from Eq.(\ref{defi}) the charge density for the unpolarized $N\to\Delta$ transition can be written as \cite{vande}
\be
\rho_0^{N\Delta}(b)&=&\int \frac{d^2\Delta_{\perp}}{(2\pi)^2}e^{i\Delta_{\perp}. b_{\perp}}G^+_{+(1/2)+(1/2)}(Q^2)\nonumber\\
&=&\int_0^\infty \frac{dQ}{2\pi}QJ_0(bQ)G^+_{+(1/2)+(1/2)}(Q^2),\label{rho_0}
\ee
where the transverse impact parameter $b=|b_{\perp}|$ and $J_n$ indicates the cylindrical Bessel function of order $n$. $G^+_{+(1/2)+(1/2)}$ is the helicity conserving $N\to\Delta$ form factor and can be expressed in terms of $G_M^*$, $G_E^*$ and $G_C^*$ as \cite{vande,tiator2}
\be
&&G^+_{+(1/2)+(1/2)}(Q^2)=I\frac{(M_N+M_{\Delta})}{M_NQ_+^2}\sqrt{\frac{3}{2}}\bigg(-\frac{Q^2}{4}\bigg)\nonumber\\
&\times&\bigg\{G_M^*+G_E^*\frac{3}{Q_-^2}\Big[(3M_{\Delta}+M_N)(M_{\Delta}-M_N)-Q^2\Big]\nonumber\\
&+&2G_C^*\Big[-\frac{(M_{\Delta}-M_N)}{M_N}+3\frac{Q^2}{Q_-^2}\Big]\bigg\},
\ee
where the isospin factor $I=\sqrt{2/3}$ for the nucleon to $\Delta(1232)$ transition.

The unpolarized charge density leads to the monopole pattern only. To get  information  about the quadrupole moments of the nucleon and $\Delta$ states,  we need to consider the charge densities for transversely polarized nucleon and $\Delta(1232)$. 
The charge density for transversely polarized nucleon and $\Delta$ is given by \cite{vande}
\be
\rho_T^{N\Delta}(b)&=&\int_0^\infty \frac{dQ}{2\pi}\frac{Q}{2}\Big[J_0(bQ)G^+_{+(1/2)+(1/2)}\nonumber\\
&+&\sin(\phi_b-\phi_s)J_1(bQ)\Big\{\sqrt{3}G^+_{+(3/2)+(1/2)}\nonumber\\
&+&G^+_{+(1/2)-(1/2)}\Big\}\nonumber\\
&-&\cos(\phi_b-\phi_s)J_2(bQ)\sqrt{3}G^+_{+(3/2)-(1/2)}\Big].\label{rho_TT}
\ee
\begin{figure*}[htbp]
\begin{minipage}[c]{0.98\textwidth}
\small{(a)}
\includegraphics[width=7.9cm,clip]{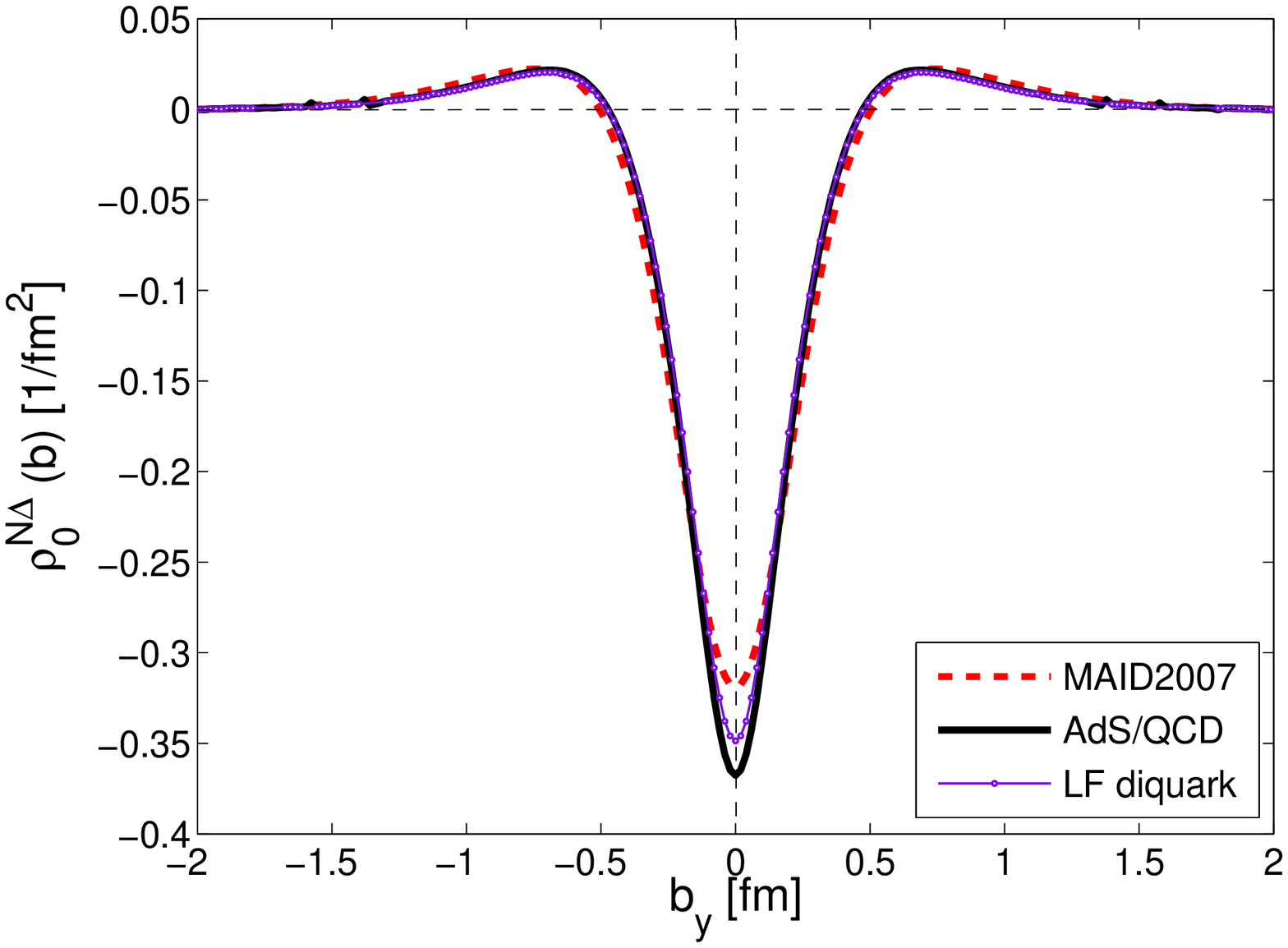}
\hspace{0.1cm}%
\small{(b)}\includegraphics[width=7.9cm,clip]{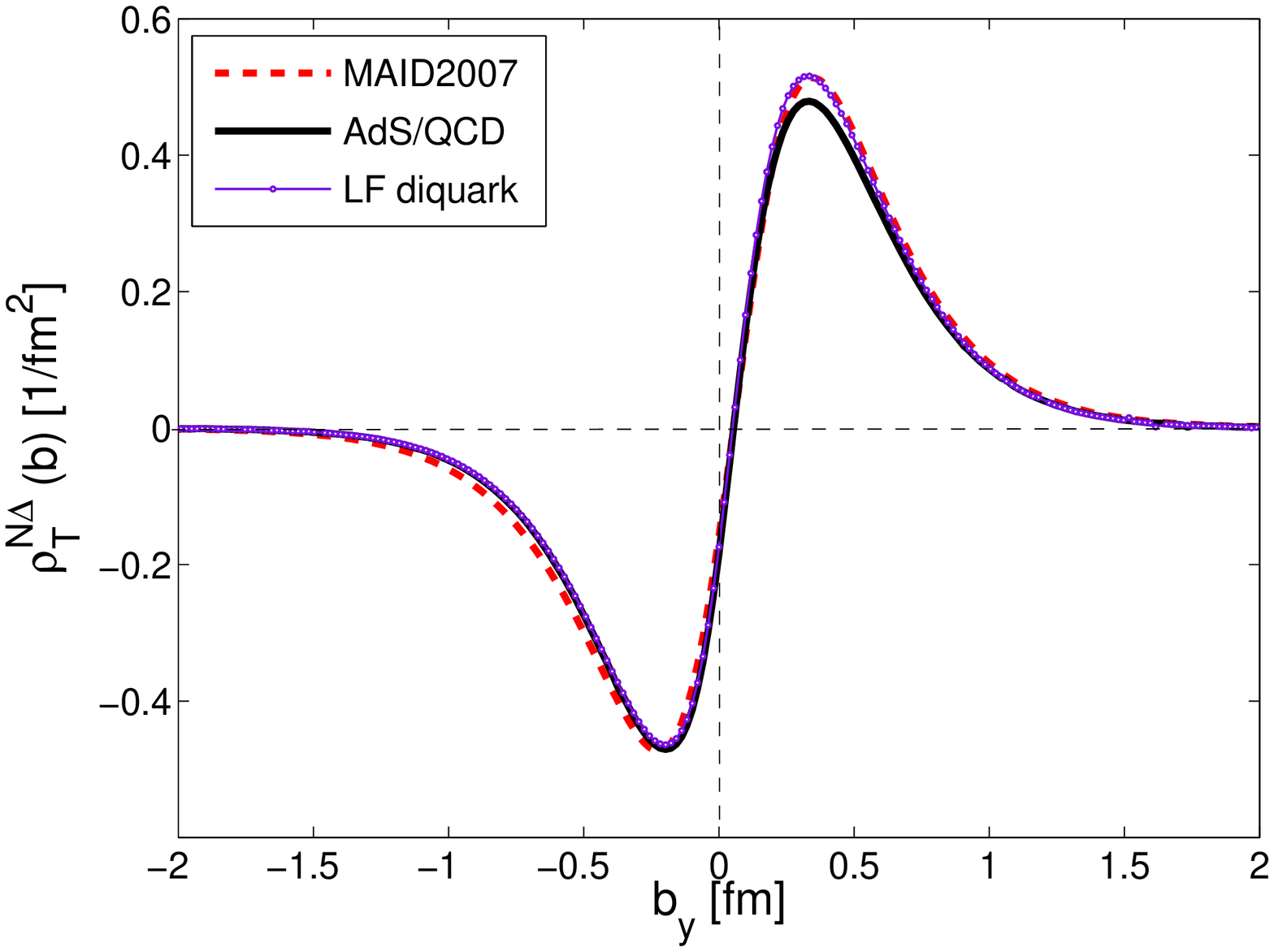}
\end{minipage}
\caption{\label{RHO_0}(Color online) The $N \to \Delta$ transition charge densities (a) unpolarized (b) both $N$ and $\Delta$ are transversely polarized along $x$-direction. The red dash line represents MAID2007 \cite{maid}.}
\end{figure*}
\begin{figure*}[htbp]
\begin{minipage}[c]{0.98\textwidth}
\small{(a)}
\includegraphics[width=7.9cm,clip]{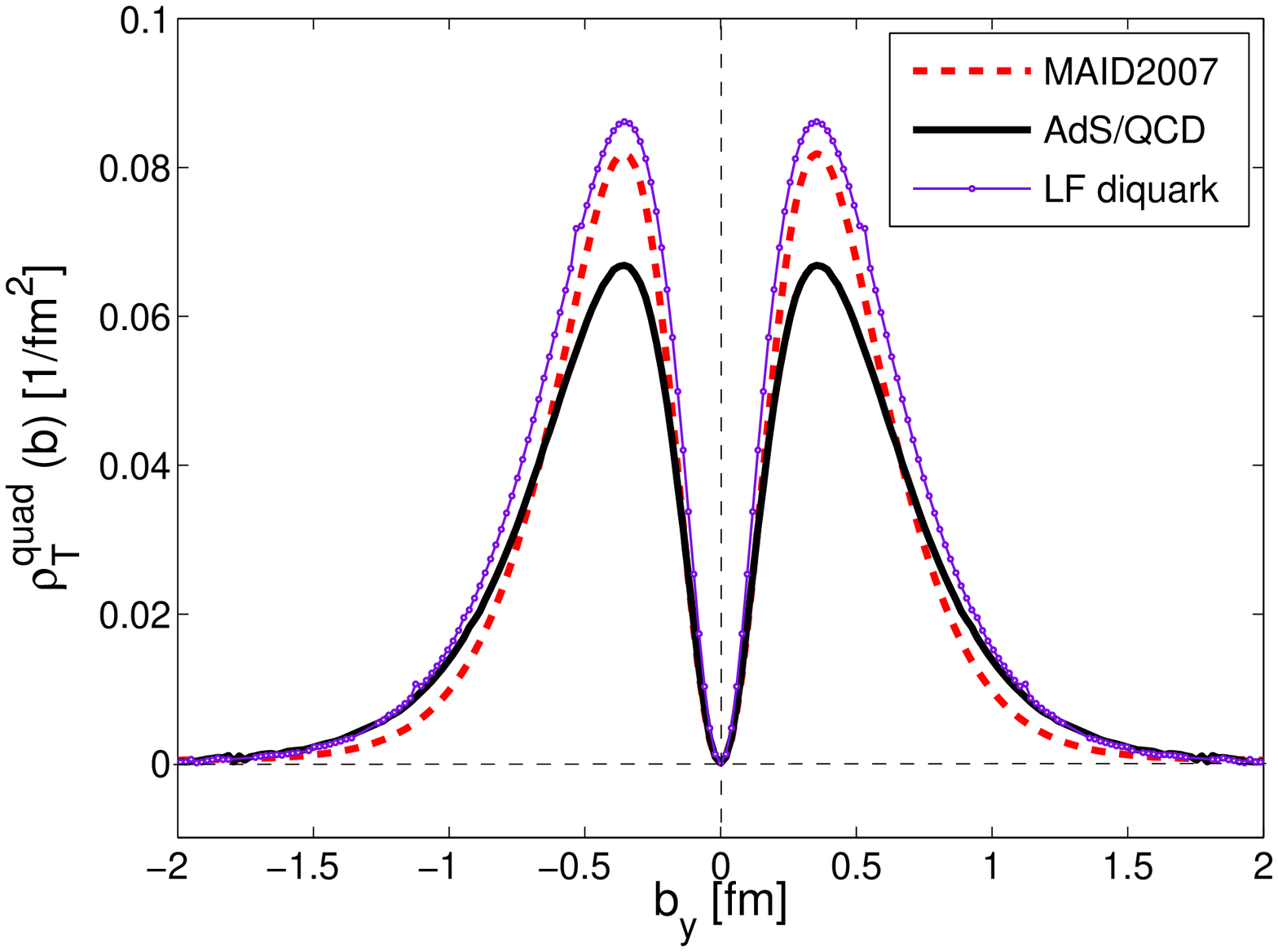}
\hspace{0.1cm}%
\small{(b)}\includegraphics[width=7.9cm,clip]{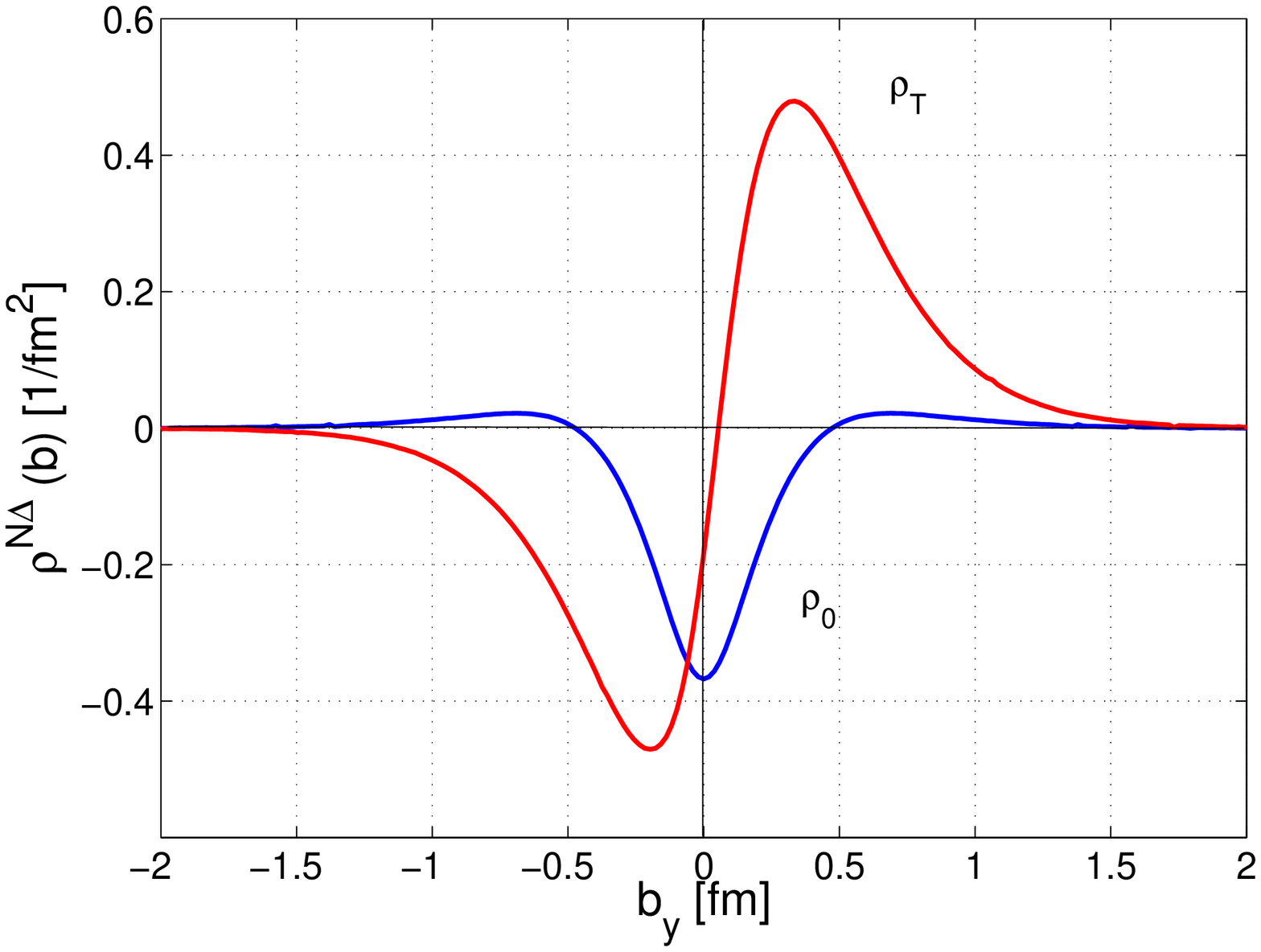}
\end{minipage}
\caption{\label{RHO_compare}(Color online) The $N\to\Delta$ transition charge densities (a) quadrupole contribution to the transversely polarized charge density, (b) comparison of unpolarized and transversely polarized densities calculated in AdS/QCD .} 
\end{figure*}
\begin{figure*}[htbp]
\begin{minipage}[c]{0.98\textwidth}
\includegraphics[width=5.45cm]{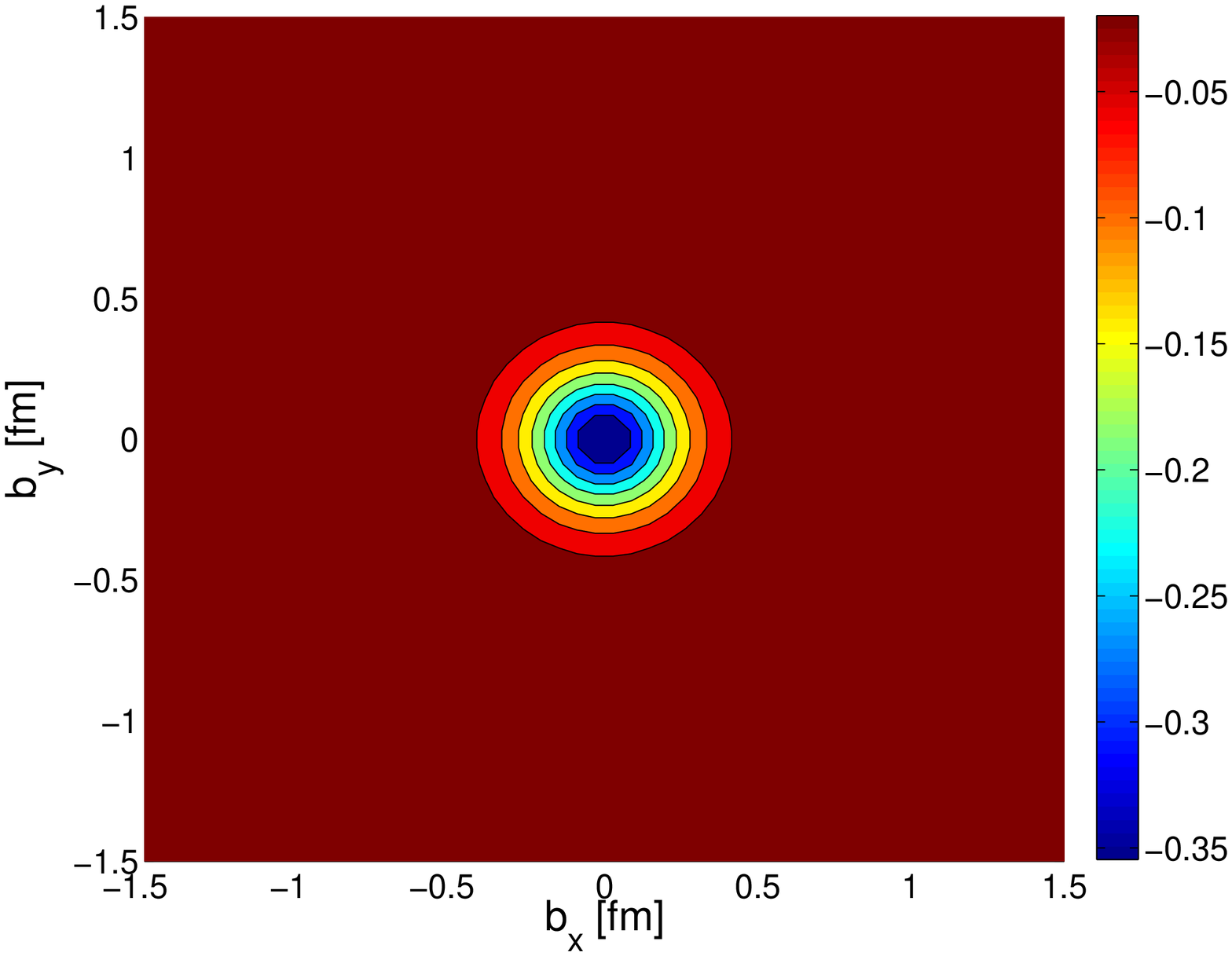}
\includegraphics[width=5.45cm]{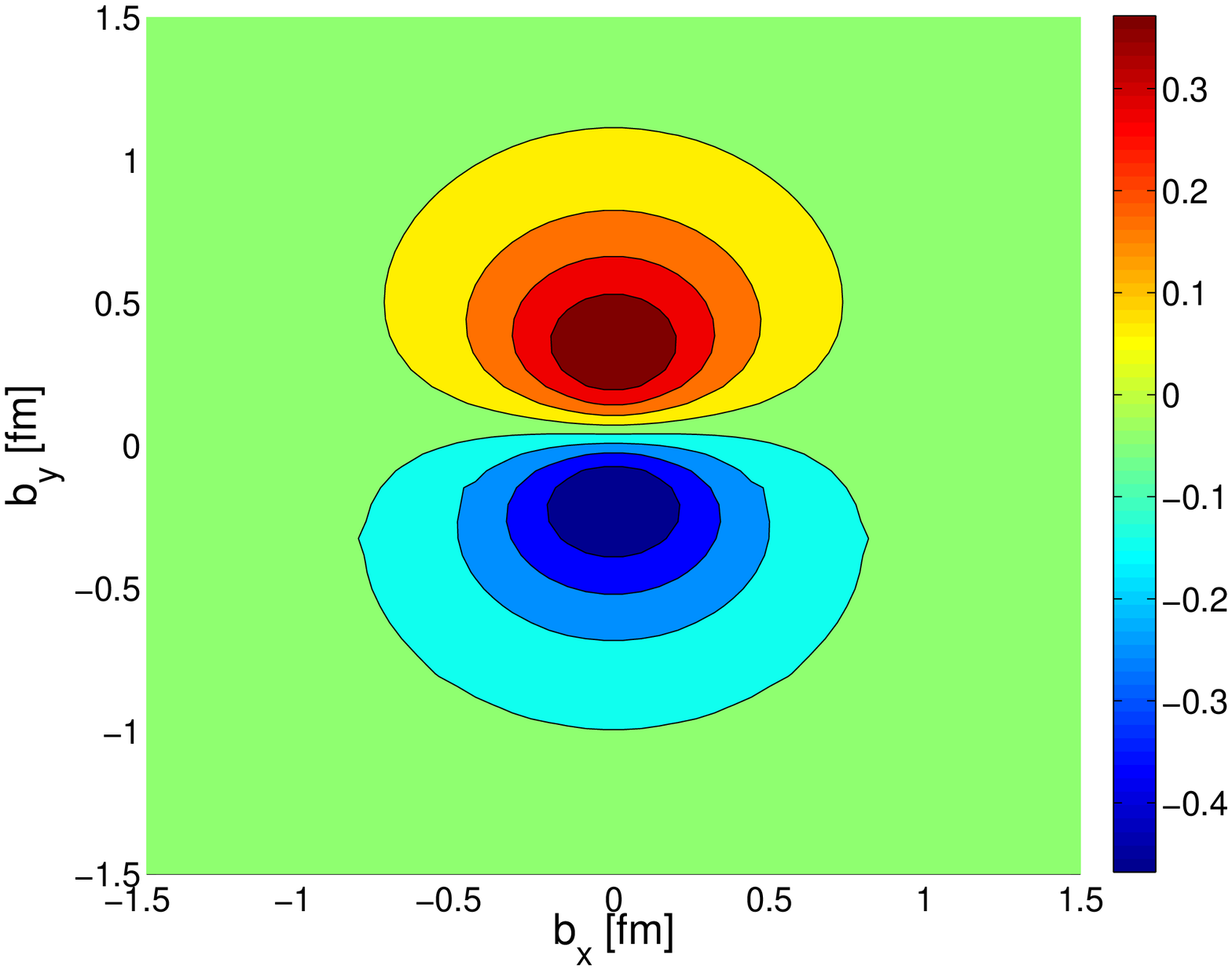}
\includegraphics[width=5.45cm]{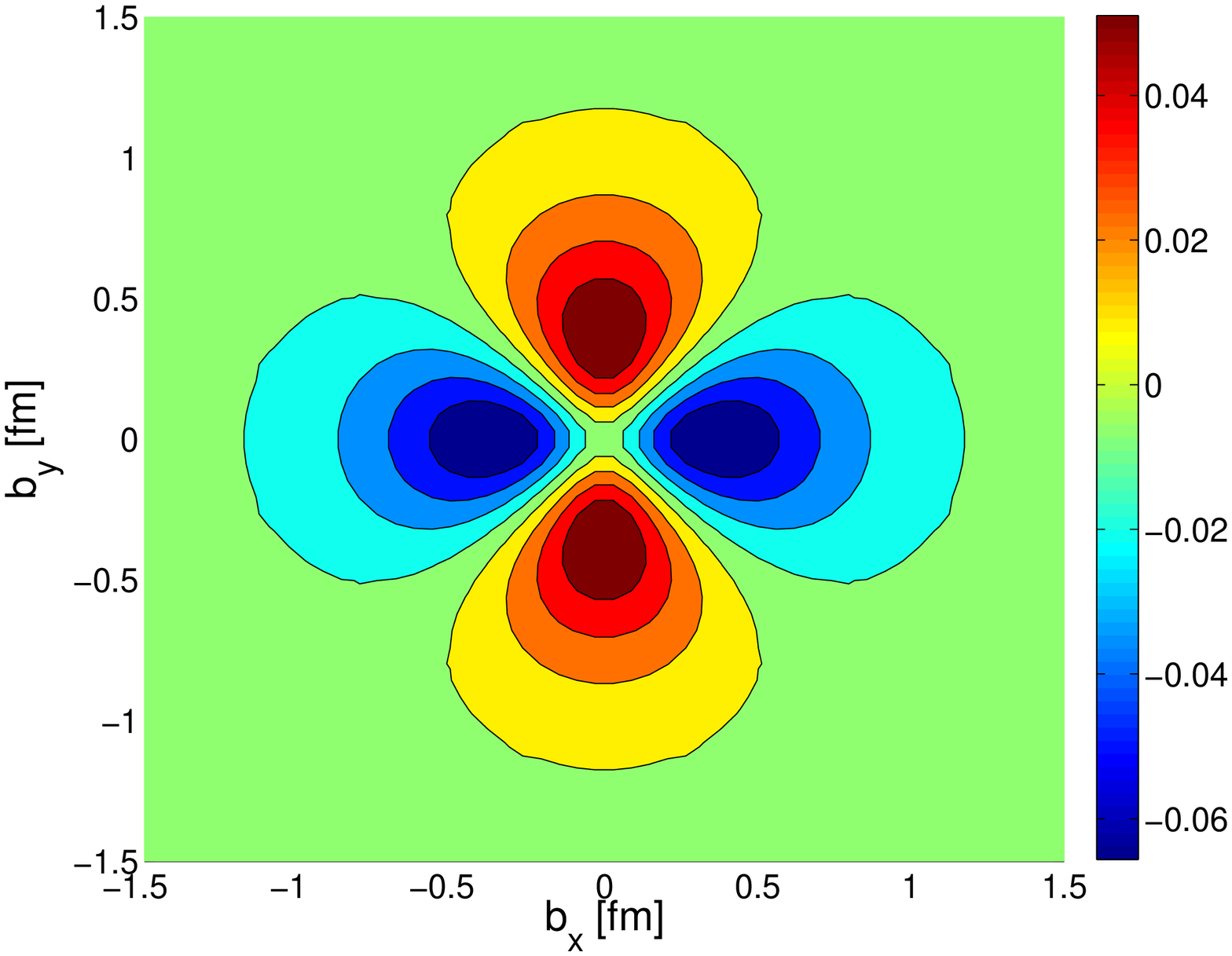}
\end{minipage}
\caption{\label{top}(Color online) Top view of the $N\to\Delta$ transverse transition charge densities, left: when both $N$ and $\Delta$ are unpolarized i.e $\rho_0^{N\Delta}$; middle: both are polarized along $x$ axis, $\rho_T^{N\Delta}$; and right: quadrupole contribution to $\rho_T^{N\Delta}$. All are evaluated in soft-wall AdS/QCD model.} 
\end{figure*}
It can be noticed in Eq.(\ref{rho_TT}) that $\rho_T^{N\Delta}$  is a linear combination of unpolarized helicity conserving  transition charge density together with two other independent components. The second term involves  one unit of light-front helicity flip($1/2 \to 3/2$ or $-1/2 \to 1/2$) nucleon to $\Delta$ transition form factor which gives a dipole field pattern in the charge density. The last term,  involves the form factor with two unit of light-front helicity flip($-1/2 \to 3/2$)  and corresponds to a quadrupole field pattern in the charge density. One writes the helicity flip form factors in terms of $G_M^*$, $G_E^*$ and $G_C^*$ as \cite{vande,tiator2}
\be
&&\sqrt{3}G^+_{+(3/2)+(1/2)}+G^+_{+(1/2)-(1/2)}\nonumber\\
&=&I\frac{(M_N+M_{\Delta})}{M_NQ_+^2}\sqrt{\frac{3}{2}}Q\bigg\{G_M^*(M_{\Delta}+M_N)+G_C^*\frac{Q^2}{2M_{\Delta}}\bigg\},\nonumber\\
&&\\
&&G^+_{+(3/2)-(1/2)}\nonumber\\
&=&I\frac{(M_N+M_{\Delta})}{M_NQ_+^2}\frac{3}{4\sqrt{2}}Q^2\bigg\{G_M^*\nonumber\\
&+&G_E^*\bigg[1-\frac{4M_{\Delta}(M_{\Delta}-M_N)}{Q_-^2}\bigg]-G_C^*\frac{2Q^2}{Q_-^2}\bigg\}.
\ee

Without loss of generality, we take the polarization of both the nucleon and $\Delta$ along $x$-axis ie., $\phi_s=0$. We show the  charge density for unpolarized nucleon to $\Delta$ transition in Fig.\ref{RHO_0}(a) using the transition form factor obtained in both AdS/QCD and quark-diquark model. The similar plot for transversely polarized nucleon and $\Delta$ has been shown in Fig.\ref{RHO_0}(b).  The consequences are compared with the results shown in Ref. \cite{vande} which were evaluated using the transition form factor from MAID2007. The unpolarized charge density shows a behavior having negatively charged core surrounded by a ring of positive charge density for $b\geq 0.5$. In both cases, the predictions of AdS/QCD and quark-diquark model are in excellent agreement with the MAID2007 parameterization except the fact that AdS/QCD gives slightly more negative value for unpolarized density at $b=0$. The quadrupole contribution to the transversely polarized density has been shown in Fig.\ref{RHO_compare}(a) 
whereas we provide a comparison of the unpolarized density $\rho_0^{N\Delta}$ with the transversely polarized density $\rho_T^{N\Delta}$ evaluated in AdS/QCD in Fig.\ref{RHO_compare}(b).
In Fig.\ref{top}, we have shown a top view plot of three dimensional transition charge densities in the transverse plane calculated in AdS/QCD for unpolarized  nucleon and $\Delta(1232)$ (left panel), as well as for both the nucleon and $\Delta$ are polarized along $x$-direction (middle panel). One notices that the unpolarized density is axially symmetric whereas the transversely polarized density shows a dipolar pattern. These behaviors are very similar to neutron charge densities observed in \cite{vande,CM3}. The dipolar pattern comes in $\rho_T^{N\Delta}$ due to large anomalous magnetic moment coming from the second term of the Eq.(\ref{rho_TT}) which produces an induced electric dipole moment in $y$-direction. The top view plot of quadrupole contribution to $\rho_T^{N\Delta}$ has been shown in the right panel of Fig.\ref{top}. One can also notice that the contribution of the quadrupole term to the deformation of the charge density is comparatively very weak, thus the distorted density effectively 
exhibits 
the dipolar pattern.  Fig.\ref{top} shows the relative strengths of the monopole, dipole and quadrupole contributions to $\rho_T^{N\Delta}$. 
\section{\bf Summary}\label{concl}
In this work, we have presented a comparative study of the nucleon to $\Delta$ transition form factors in terms of nucleon electromagnetic form factors in the framework of a soft-wall AdS/QCD model and a light-front quark-diquark model.  We have used the large $N_c$ formulas to evaluate the transition form factors from the nucleon electromagnetic form factors. We have compared the results with the available experimental data as well as with the standard parameterization, MAID2007. $G^*_M(Q^2)$ in both models are in more or less agreement with experimental data and MAID2007. It has also been found that the AdS/QCD predictions for transition ratios are in good agreement with the experimental data and better than that of the quark-diquark model and MAID2007. Further, we have investigated the transition charge densities in the transverse plane by taking the Fourier transform of the transition form factors. Both the unpolarized nucleon and $\Delta$ and the transversely polarized nucleon and $\Delta$ are 
considered here. 
The densities in both AdS/QCD and quark-diquark models are consistent with the results of the Ref. \cite{vande}. Though AdS/QCD gives only semiclassical approximation of QCD, it reproduces the $N\to \Delta$ transition data very well. 
The unpolarized density is axially symmetric and it gives only monopole pattern but the transversely polarized density provides all the monopole, dipole and quadrupole patterns. The quadrupole contribution to $\rho_T^{N\Delta}$ is comparatively small and effectively the transversely polarized density shows a dipolar pattern.



\end{document}